\documentclass[runningheads]{llncs}
\usepackage{graphicx,amsmath,todonotes,hyperref,subcaption,cleveref,tabularx,algorithm,algorithmicx,algpseudocode}
\usepackage{tikz}

\newcommand{\ExternalLink}{%
    \tikz[x=1.2ex, y=1.2ex, baseline=-0.05ex]{%
        \begin{scope}[x=1ex, y=1ex]
            \clip (-0.1,-0.1) 
                --++ (-0, 1.2) 
                --++ (0.6, 0) 
                --++ (0, -0.6) 
                --++ (0.6, 0) 
                --++ (0, -1);
            \path[draw, 
                line width = 0.5, 
                rounded corners=0.5] 
                (0,0) rectangle (1,1);
        \end{scope}
        \path[draw, line width = 0.5] (0.5, 0.5) 
            -- (1, 1);
        \path[draw, line width = 0.5] (0.6, 1) 
            -- (1, 1) -- (1, 0.6);
        }
    }

\begin{document}
\title{Observations on Transitioning to Teaching Computer Science Online}
\author{Mehrnoosh Askarpour}
\authorrunning{M. Askarpour}
\institute{Computing and Software, McMaster University, Canada\\
\email{askarpom@mcmaster.ca}
}
\maketitle             
\begin{abstract}
The hit of the COVID-19 pandemic has hugely affected higher education in the world, and as a result, most of the physical classes have been (partially) replaced by online teaching platforms. This transition is challenging even for experienced software engineering instructors, as they were pushed to break their habits and tricks developed over the years
. This paper is an experience report in teaching an undergraduate course (revolving theoretical computer science topics) for the first time in an online format, and some observations and ideas of how to engage students during online lectures.
\end{abstract}


\section{Introduction}
Teaching the basic and theoretical concepts of computer science, such as Automata Theory, has always been an essential part of the computer science and engineering curriculum. Given that there are several resources out there for students to refer to, books such as \cite{kozen2012automata,mandrioli1987theoretical} and an extensive amount of online material, instructors have to organize their classes not only to explain the material but to engage the students and promote active learning. In regular classes, this could be done by asking questions, inviting volunteers to the board, evaluating student's feedback by directly interacting with them in the class, in addition to assignments, exercises, etc.

However, we have been far from a regular situation since the hit of COVID-19 in March 2020, when most of the universities limited their classes in the presence and switched to online/hybrid teaching. The situation seemed to be temporary, and many countries already have had thorough vaccination programs. However, many countries with a considerable share in exporting international students have yet to vaccinate their population. This matter and the huge environmental impact of limited travelling (e.g., international flights or even local commute) and the consequent save in time and cost could contribute to the idea of hybrid classes at universities for good.

This transition has obvious impacts on the teaching philosophy of many software and engineering instructors and urges them to try new tools and ideas. In particular, it is challenging for young and inexperienced instructors to cope with the peculiarities of teaching in this new situation. If engaging students and verifying their performance has always been an issue for theoretical computer science courses, one can imagine it is even more challenging for new instructors in regular classes, let alone online format.

I instructed an undergraduate course on the foundations of computer science for the first time, entitled ``Discrete Mathematics with Applications II", in winter 2021 at the Department of Computing and Software of McMaster University, which was also my first teaching experience online (read about my previous experience in \cite{10.1007/978-3-030-57663-9_1}). This paper discusses my observations, the main challenges of building a meaningful interaction with students, and a few suggestions for similar courses that hopefully would be useful for other young instructors.

The rest of this paper is structured as follows: \cref{sec:course} explains the syllabus and structure of the course I use as a running example throughout the paper, \cref{sec:challenges} lists a few of my observations and challenges in interacting with students, \cref{sec:sug} provides some suggestions, and \cref{sec:con} concludes.
\section{Structure of the Course}
\label{sec:course}
The course I instructed last semester, entitled ``Discrete Mathematics with Applications II", is presented in \cref{tab:mac-profile}. It is a mandatory course with three units in the undergraduate computer science program at McMaster University. It was a large class with nearly 200 students, mainly computer science students, except for approximately 36\% of the class attending other engineering programs. I delivered the course via three main platforms:
\begin{itemize}
    \item \textbf{Youtube}: Prerecorded videos of the main concepts of each week were posted here to be watched by students before discussion sessions.
    \item \textbf{Microsoft Teams}: Live sessions discussing the material and Q\&As were held where the students were supposed to attend twice a week.
    \item \textbf{Avenue to Learn\footnote{\url{https://mi.mcmaster.ca/avenuetolearn/}}}: Lecture slides, assignments, and anything else required to be passed on to the students were posted on this platform which is the primary learning, course delivery and assessment platform at McMaster University.
\end{itemize}

Organizing a large class online might seem much easier and less intimidating for a young instructor at first glance; however, there are also inevitable new challenges. Perhaps the most important issue is the loss of direct and face-to-face contact with students that could be beneficial both for students and instructors. The following section lists the issues that I encountered throughout this course and the solutions I chose for facing them. 

\begin{table}[t]
\centering
\begin{tabularx}{\textwidth}{l|X}
\hline
Title & Discrete Mathematics with Applications II\\
\hline
Students Level & Second Year of B.Sc. in Computer Science\\ \hline
Number of Students & 143 computer science students + 52 other engineering students (excluding software and computer engineering students) \\ \hline
Course Type & Mandatory \\ \hline
Required Background & Discrete Mathematics with Applications I \href{http://www.cas.mcmaster.ca/~kahl/CS1FC3/2013/}{\ExternalLink}
(Functions, relations and sets; the language of predicate logic, propositional logic; proof techniques, counting principles; induction and recursion, discrete probabilities, graphs, and their application to computing).
\\\hline
The Syllabus & Mathematical proof, \\
& Recursion and induction,\\
& Predicate logic,\\
& Finite automata and regular expressions,\\
& Push-down automata and context-free languages,\\
& Turing machines and computability.\\
\hline
Text Book& Dexter C. Kozen, Automata and Computability, Springer, 1997.\\
\hline
Evaluation &  10 assignments throughout the course and three exams (2 midterms and a final exam).\\
\hline
\end{tabularx}
\vspace{.5cm}
\caption{The profile of Discrete Mathematics with Applications II course taught at McMaster University in Winter 2021.}
\label{tab:mac-profile}
\end{table}

\section{Challenges and Observations}
\label{sec:challenges}
Like I said earlier, this was the first time I was fully responsible for a course that happened to be a very large undergraduate class. Thus, many of my remarks could be interpreted in a general sense about teaching and not only about transitioning to online teaching. Below I discuss the most challenging observations: 
\begin{itemize}
    \item \textbf{The loss of face-to-face communication with the students}: Naturally, direct contact between the students and their instructors is compromised due to the online teaching format. At first, one might think that works in favour of an inexperienced instructor because they will not be intimidated by the size of the class. However, building a productive relationship with students 
    is more difficult with the new format. The online format gives the students some sense of anonymity; So, they might not follow the same moral code they usually follow in the class (e.g., being online but doing something else, making inappropriate comments or having side conversations in the chat). Moreover, since some universities impose recording the online discussion sessions, students developed a tendency for skipping live sessions and watch the recordings later (e.g., close to the exam which leaves them very little time to go through all the videos precisely) which consequently negatively affects their participation in the course and the quality of live classes. 
    \item \textbf{International students}: Since the beginning of the pandemic, most of the international flights have been cancelled which led to some of the students not being able to fly back to where their universities are located. As a result, we had participants from several counties in class (e.g. China, India, Pakistan, Vietnam, Canada) who lived in different time zones. It was sometimes very difficult for them to follow the scheduled classes which were based on CET. For example, the live sessions were scheduled for 14:00 CET which was 2:00 A.M. in China.
    This issue needed extra attention from the instructors to accommodate these students and give them an equal chance of accessing the class material and discussions. 

    \item \textbf{Evaluating the participant and engagement of students in online classes}:  A challenging issue is to encourage students to participate in classes as seriously as in the physical classes. One feature of Teams that helps is the participation list downloadable after each call, but it cannot be trusted because people might just show up online while doing other things, or not even being by their PCs. So, instructors should come up with mechanisms to boost student's willing participation.
    \item \textbf{Webcam on or off}: Many think that letting the students leave their webcam off will give them more freedom and tranquillity; So, the most common policy is not to force anyone to keep their camera on. I tried the webcam-off approach and strongly felt the loss of direct contact and communication. When everyone's webcams and microphones are off, it is difficult to have any feedback from the students about the speed of the lecture or realize if they are following.\\
    Besides, using the chat as a feedback provider or a means to ask questions during the lecture slows down the rhythm of the course noticeably and sometimes makes the instructor lose their train of thought. 
     \item \textbf{Prohibit cheating}: Organizing exams in a way to limit students cheating is very difficult for online teaching but also essential not to worsen the quality of education in general. It is important to note that in remote learning, a closed book exam is almost impossible because there is no way to surveillance the students during the exams. Some universities started to use camera-on policies during the exams which led to some complaints about the violation of privacy from the students' side\footnote{\href{https://www.insidehighered.com/advice/2021/03/03/why-its-wrong-require-students-keep-their-cameras-online-classes-opinion}{insidehighered.com/camera-policy-for-online-schools}}. Thus, the safest solution would be to suppose that the exam will be open book, then design the format and questions accordingly.\\
     Additionally, students might communicate with others during the exam and try to get help. They also might use online material to find the answers to questions of the exams or assignments.
    \item \textbf{Assignments}: Assignments are one of the best ways to evaluate the state of the class and could potentially contribute to the final evaluation of the students. They are even more important for online classes because this sense of having everything recorded and online entices students to procrastinate to study until the end of the semester. 
    However, this part of transitioning to online teaching is not very difficult, since we have been very well prepared for it. 
    Most of the universities already had  online submission platforms for assignments and projects from long ago. Additionally, many instructors used free platforms such as Github and bitbucket. However, extra attention is required for avoiding cheating (see the fifth item of this list) and assigning deadlines for the sake of international students (see the second item of this list).
    \item \textbf{Replacing group works}: Activity-based learning and teamwork proved to be very effective for teaching Automata Theory~\cite{6306722}. However, this is hugely affected by the switch to the online teaching format. Students will not meet each other and sometimes are even in different towns. First-year students did not even have the chance to meet their classmates and get to know them. So, it is important to give them a structure to communicate, collaborate and learn from each other. This is possible via Slack, Teams, and discussion forums of Avenue to learn, and probably many more tools out there~\cite{tools1,tools2,tools4}. Instructors could create different channels/threads for different topics of the course, exams, and assignments so that students can exchange ideas about them in an easier and more organized way.
\end{itemize}

\section{Suggestions}
\label{sec:sug}
This section provides my experience in dealing with the challenges introduced in the previous section, in addition to some general observations.
\begin{itemize}
 \item \textbf{One-to-one communication}: We have to take into account that the transition to online teaching is new to students too, especially those who are in their first two years of the university who are relatively young and just getting used to the university format. Hence, it is very important to remain available and reachable to them and make them feel comfortable to approach you and ask questions. When offices were still open, students would just come during office hours and ask questions. The same thing could be done online by scheduling weekly office hours and encouraging students to show up and ask questions one on one. These hours should be selected considering the limitations of the international students in terms of their time zones.
  \item \textbf{(Pre-) Recording classes and multiple exam slots}: To solve the problem of different time zones, the class topics were covered in Youtube videos, no longer than 30 minutes each. The videos were being uploaded twice a week and the students were asked to watch them before coming to the live sessions (twice per week). This would allow students attending the course from anywhere to watch them at their convenience and not miss the classes due to bad timing (the next item explains how to push the students not to prolong this to the end of the semester). Pre-recorded videos allow students to review the material multiple times if they have doubts and questions.
  Moreover, the instructor could build some estimation of the engagement of the students by the number of views or comments.
  
 \item \textbf{Daily quizzes to monitor student participation.}: One way to solve the problem of monitoring the attendance of students to the class, is to ask questions from students. However, students do not love being put on the spot. Besides, it might be very time consuming for a large class. So, I found short quizzes each session a better solution. They could be managed through an assessment platform like Avenue to learn or any other available tool~\cite{tools3}. These quizzes should be taken at random times during each session. Experience shows that if you always do it at the beginning or the end of each session, they will not resolve the problem of inattentive participation. It would also be very interesting to prepare the quiz question in advance but related to one of the examples you will show in the class so the students will follow the conversation more eagerly.\\
 However, international students were allowed to skip one quiz per week without affecting their final evaluation. The live sessions (which revolved around live Q\&A discussions about the topics covered in the pre-recorded videos) are recorded as well so students can reiterate the solved exercises in class several times as they need.

 \item \textbf{Organizing online exams}: The exams should be organized in at least two sessions to accommodate international students. No one likes to take an exam at 3:00 A.M. The format and timebound of the exam should limit the possibility of students exchanging the answers or get advice from others. We went for multiple choice and short answer questions so that correcting all the exams could be automated to a good extent. One might fairly argue that these types of questions are not suited for the syllabus of this course. This is indeed true; for example, instead of asking the students to construct an automaton for a certain language, they were given a few options to choose the correct one from. This could make each question a bit easier at the first glance but could be balanced out by proper timing and clever questions. Moreover, asking students to draw automata and then upload them would complicate the process of evaluation, in addition to many problems that always happen during the exams by the students in uploading files correctly in the right file format by the students.\\
 To stop students from cheating, it is essential to have fresh questions for the exam and not to re-use the questions of earlier exams or anything whose answer could be easily found by a Google search.\\
 Furthermore, most of the exam organizing platforms, including Avenue to learn and Moodle\footnote{\url{https://moodle.org/}}, do not allow to navigate questions of an exam in any arbitrary order. One can limit the navigation to only moving forward and while setting up the exam so students won't be able to navigate back and forth; Therefore, if each student gets a different order of questions (randomizing questions), the possibility of them exchanging answers to questions will decrease.
 
 Another effective solution to this problem is to encourage students to read the student conduct code document and talk about the definition of academic dishonesty and its instances in class with students in a friendly yet informative manner.
 
 \item \textbf{It is not the end of the world if you make mistakes}: Imposter syndrome has been a common issue among many colleagues that I had along with my academic career. 
  Additionally, some studies showed that students in general take it harder on female instructors~\cite{KENG2020101889} which could lead to higher stress levels among them.\\
  To address these struggles, we need to create and maintain strong networks (e.g. FMTea, Frontiers in Software Engineering Education) where instructors share experience, tips, and ideas and encourage the community to avoid double standards that only cause stress and lower our performance.
  
  \item \textbf{Using off-the-shelf-tools}: In addition to communication tools discussed in the previous section, using tools to teach Automata Theory have been long used as a way to make the topic more interesting~\cite{DBLP:conf/sigcse/RodgerBLPPST97,DBLP:journals/bjet/ChakrabortySK12,7344185}. It would also help the students better prepare their assignments which in turn eases their evaluation.
  \item \textbf{Keep updated}: There are tons of shared experiences online (e.g., \cite{share1,share2,share3}) about this transition to hybrid or online teaching which are very helpful. It might be very beneficial to read these experience reports and use their advice and suggestions. 
\end{itemize}

\section{Conclusions}
\label{sec:con}

This paper is an experience report of teaching an undergrad course by myself. The course, which covered the fundamentals of theoretical computer science, had to be fully transformed to fit in the online teaching format. I had plenty of experience in teaching assistant activities earlier but this was my first time taking the full responsibility of an undergraduate course. Besides the usual perks of teaching, this paper summarizes what I observed as challenges of online teaching for such classes. 

In a nutshell, the new online teaching format requires the instructors to dedicate some time to make up for the lost in person contact with students by online office hours, take inclusiveness serious and be sensitive to the difficulties of international students, organize exams suitable for the online format, try to stay updated about teaching tools and methods, and finally exchange experience with other instructors.

\section*{Acknowledgement}
The reported course has been delivered in parallel to the same course for software engineering students, which has been organized and held by Prof. William Farmer. Many of the ideas I used throughout my course has been suggested to me by him which helped me to structure this course better and easier. I would like to thank him for his continuous support and generosity in sharing his valuable knowledge and experience with me.

\bibliographystyle{splncs04}
\bibliography{bib}

\end{document}